\documentclass[epj,final]{svjour}
\usepackage{psfig}

\begin{document}

\title{Raman cross section of spin ladders}

\author{Christoph Jurecka, Verena Gr\"utzun, Andreas Friedrich
\and Wolfram Brenig}                     
\authorrunning{C. Jurecka, et al.}

\institute{Institut f\"ur Theoretische Physik, Technische Universit\"at
Braunschweig, 38106 Braunschweig, Germany}

\date{Received: date / Revised version: date}

\abstract{
We demonstrate that a two-triplet resonance strongly renormalizes the
Raman spectrum of two-leg spin-ladders and moreover suggest this to be
the origin of the asymmetry of the magnetic Raman continuum observed in
CaV$_2$O$_5$.
} 

\maketitle
  
\begin{figure}
\psfig{file=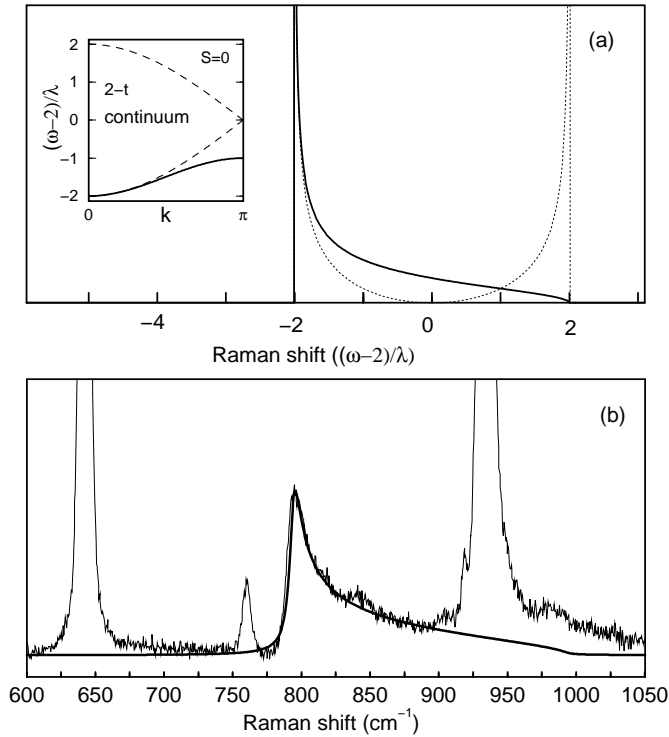,width=\linewidth,angle=-90}
\caption{(a) Solid line: Raman intensity (\ref{Iw}). Dotted line:
bare Raman intensity neglecting two-triplet interactions.
Inset: $S=0$ two-triplet spectrum of spin ladders. (b) Thin solid line:
experimental Raman spectrum (after [2]). Thick solid line:
fit of theory to experiment, see text.}
\label{fig1}
\end{figure}

Magnetic Raman scattering is a powerful tool to investigate the total
spin-zero excitations near zero momentum in low-dimensional quantum-spin
systems~\cite{Lemmens99}. In a recent Raman scattering study by
Konstantinovi\'c and collaborators~\cite{konstan00} a strongly asymmetric
magnetic continuum, see fig.~\ref{fig1}(b), has been observed in the
spin-ladder compound CaV$_2$O$_5$. It has been realized by the authors of
this study that the continuum defies an interpretation in terms of
non-interacting two-triplet excitations as given in ref.~\cite{orignac00}.
The latter would imply {\em two} van-Hove-type intensity maxima, one at
the lower and one at the upper edge of the two-triplet continuum.
Noteworthy, the magnetic Raman intensity for the two-leg spin-ladder has
been evaluated also by exact diagonalization (ED)~\cite{natsume98}. Within
the limitations of finite system analysis the ED results are consistent
with the observed intensity if the intra-rung coupling on the ladder is
assumed to be strong in CaV$_2$O$_5$, moreover, the ED is incompatible
with the non-interacting spectra of ref.~\cite{orignac00}. While this
clearly emphasizes the relevance of interaction effects, it is unfortunate
that no simple physical picture can be extracted from the ED data to allow
for a direct interpretation of the measured Raman spectrum.

In this brief note we clarify that the physical origin of the asymmetric
Raman continuum of two-leg spin-ladders is a {\em two-triplet bound state}
of total spin zero which merges with the two-triplet continuum at small
wave vector to form a resonance. Our analysis is focussed on the limit of
strong intra-rung coupling which is one likely scenario also for the
magnetic properties of CaV$_2$O$_5$~\cite{konstan00,Johnston00}. In this
limit we can profit from an exact evaluation of the two-triplet propagator
which has been carried out including all two-triplet interactions in a
different study of phonon-assisted two-triplet optical absorption (PTA) of
spin-ladders~\cite{jurecka00}.

The Hamiltonian of the two-leg spin-ladder reads
\begin{equation}
\label{e1}
H=\sum_{l,\alpha} [S^\alpha_{1 l} S^\alpha_{2 l}
+\lambda (S^\alpha_{1 l} S^\alpha_{1 l+1}+
S^\alpha_{2 l} S^\alpha_{2 l+1})]
\end{equation}
where $S^\alpha_{\mu l}$ with $\alpha=x,y,z$ is a spin-1/2 operator on
site $l$ of leg $\mu$ and H is measured in units of $J_\perp$ with
$\lambda=J_\parallel/J_\perp$. Mag\-netic light scattering is described by
the Loudon\--Fleu\-ry vertex~\cite{fleury68}, which for the two-leg
spin-ladder is
\begin{eqnarray}
H_R=R \sum_{l,\alpha} (S^\alpha_{1 l} S^\alpha_{1 l+1}+
S^\alpha_{2 l} S^\alpha_{2 l+1})
\end{eqnarray}
where $R$ depends on the polarizations of the incident and scattered
light~\cite{freitas00}. $H_R$ is similar to the vertex for
PTA~\cite{jurecka00}, simplified however by the lack of an additional
summation over phonon coordinates.  The Raman intensity $I(\omega)$ at
zero temperature is obtained from Fermi's golden rule
\begin{eqnarray}
\label{e3}
I(\omega)&&=2\pi \sum_{f} |\langle f| H_R |0\rangle|^2
\delta(\omega-E_{f})
\\ \label{e4}
=&& -2 \;\mbox{Im}\; \sum_{q,q'} [\langle 0| H_R^\dagger
|q\rangle\langle q|\frac{1}{z-H}
|q'\rangle\langle q'| H_R^{\phantom{\dagger}}|0\rangle]
\end{eqnarray}
where $z=\omega+i0^+$. $|0\rangle$ ($|f\rangle$) are the interacting
ground (excited) states with energy 0 ($E_f$) and total momentum and spin
zero. For $\lambda\ll 1$ and following~\cite{jurecka00} $|0\rangle$ is a
product of rung-singlets and $|f\rangle$ are {\em interacting} two-triplet
excitations. Neglecting quantum fluctuations which change the number of
triplets only at $O(\lambda^2)$ the states $|f\rangle$ can be expanded in
terms of an appropriately symmetrized basis $|q\rangle$ of two-triplet
rung excitations
\begin{equation}
|q\rangle = \frac{1}{\small\sqrt{N(N-1)}} \sum_{l,m}
\mbox{sgn}(l-m)\mbox{sin}(q(l-m))|lm\rangle
\label{e5} 
\end{equation}
where $|lm\rangle=\sum_\alpha|t_{l\alpha}t_{m\alpha}\rangle/\sqrt{3}$
refers to a singlet combination of two rung-triplets created within
$|0\rangle$. The states $|q\rangle$ resemble all spin-zero two-triplet
plane-waves of zero total momentum {\em constrained} by the symmetry
$|t_{l\alpha}t_{m\beta}\rangle=|t_{m\beta}t_{l\alpha}\rangle$ and the {\em
hard-core} condition $|t_{l\alpha}t_{l\beta}\rangle=0$. The remaining
resolvent in (\ref{e4}) can be evaluated in closed form by a T-matrix
resummation (see~\cite{jurecka00})
\begin{equation}
\frac{\lambda
I(\omega)}{R^2}=\frac{3}{4}\mbox{Im}\left(\sqrt{1-\frac{4}{2+
\tilde{\omega}}}-1\right)
\label{Iw}
\end{equation}
In the limit of $\lambda\ll 1$, the intensity is a function of the
rescaled Raman-shift $\tilde{\omega}=(\omega+i0^+-2)/\lambda$ only. While
the largest energy scale, i.e. the two-triplet hard-core, is incorporated
in the states $|q\rangle$ by construction, the T-matrix resummation
accounts for both, the dispersion and the nearest-neighbor (NN) attraction
which is mediated on the two-particle level by Hamiltonian (\ref{e1}).

The thick solid line in fig.~\ref{fig1}(a) is the intensity (\ref{Iw}).
The thin solid line in fig.~\ref{fig1}(b) is the intensity measured on
CaV$_2$O$_5$ at 10K (reproduced from~\cite{konstan00}). In addition to
phonons the observed spectrum shows a strongly asymmetric line with an
onset at 795cm$^{-1}$. The thick solid line in fig.~\ref{fig1}(b) is a
comparison of (\ref{Iw}) to the experiment which results from (i) using
$\lambda=0.11$ which is consistent with~\cite{konstan00,Johnston00}, (ii)
setting $J_\perp=447$cm$^{-1}\equiv 643K$ which agrees
with~\cite{konstan00,Johnston00} and yields a continuum onset at
795cm$^{-1}$, (iii) including a broadening of $\sim$3cm$^{-1}$ by setting
$\omega+i0^+\rightarrow\omega+i0.007$ to account for instrumental
resolution, and (iv) by adjusting the arbitrary y-axis scales for a
reasonable match of the absolute intensities.  While more ambitious
fitting procedures can be envisaged the preceding is sufficient to claim
that the agreement between experiment and our theory is very good.
Moreover, we note that the solid line in fig.~\ref{fig1}(b) is consistent
with the intensity distribution obtained from ED~\cite{natsume98}.

The dotted line in~\ref{fig1}(a) depicts the bare Raman
intensity~\cite{orignac00} which results from neglecting the two-triplet
on-site hard-core as well as the NN-attraction. Displaying two van-Hove
singularities this spectrum fails to explain the observed magnetic
line-shape.

The physical origin of the asymmetric line-shape is clarified in the inset
of fig.~\ref{fig1}(a). While Raman scattering detects only zero momentum
excitations the inset reproduces the interacting two-triplet spectrum in
the spin-zero channel for $\lambda\ll 1$ over all of the Brillouin zone
from ref.~\cite{jurecka00}. Apart from the bare two-triplet continuum this
spectrum shows a bound-state induced by the two-triplet interactions which
{\em merges} with the continuum at zero momentum. This leads to a
resonance at the bottom of the continuum and to the asymmetric
redistribution of the Raman intensity. This resonance feature has to be
contrasted against Raman intensities in other low-dimensional quantum spin
systems where bound states tend to occur as sharp excitations within the
spin gap~\cite{Lemmens99}.

Finally, based on the results of high-order series
expansion~\cite{trebst00} it is tempting to speculate on the evolution of
the Raman continuum as $\lambda\rightarrow 1$. In that limit the spin-zero
bound-state merges with the continuum already at finite momentum.
Therefore, as $\lambda$ increases one might expect the resonance to shift
further into the center of the continuum. This suggests that an analysis
analogous to this work of Raman data on compounds containing
spin-ladders with $\lambda\sim 1$, 
e.g. (Ca,La)$_{14}$Cu$_{24}$O$_{41}$, should be interesting to perform.

This research was supported in part by the Deutsche Forschungsgemeinschaft
under Grant No. BR 1084/1-1 and BR 1084/1-2.


\begin{thebibliography}{}
\bibitem{Lemmens99}
P. Lemmens, M. Fischer, M. Grove, P.H.M. v. Loosdrecht, G. Els, E.
Sherman, C. Pinettes, and G. G\"untherodt, {\it Advances in Solid
State Physics Vol. 39} (Vieweg, Braunschweig, 1999).
\bibitem{konstan00}
M. J. Konstantinovi\'c, Z. V. Popovi\'c, M. Isobe, and Y. Ueda,
Phys. Rev. B {\bf 61}, 15185, 2000.
\bibitem{orignac00}
E. Orignac and R. Citro,
Phys. Rev. {\bf 62}, 8622 (2000).\\
and fig. 2 therein.
\bibitem{natsume98}
Y. Natsume, Y. Watabe, and T. Suzuki,
J. Phys. Soc. Jpn. {\bf 67}, 3314 (1998).
\bibitem{Johnston00}
D.C. Johnston, M. Troyer, S. Miyahara, D. Lidsky, K. Ueda, M. Azuma,
Z. Hiroi, M. Takano, M. Isobe, Y. Ueda, M.A. Korotin, V.I. Anisimov,
A.V. Mahajan, and L.L. Miller,
cond-mat/0001147
\bibitem{jurecka00}
C. Jurecka and W. Brenig,
Phys. Rev. B {\bf 61}, 
14307 (2000).
\bibitem{fleury68}
P. A. Fleury and R. Loudon, Phys. Rev. {\bf 166}, 514 (1968).
\bibitem{freitas00}
P. J. Freitas and R. R. P. Singh, Phys. Rev. B {\bf 62}, 14113 (2000).
\bibitem{trebst00}
S. Trebst, H. Monien, C. Hamer, Z. Weihong, and R.R.P. Singh,
Phys. Rev. Lett {\bf 85}, 4373 (2000). 
\end{thebibliography}
\end{document}